\documentclass[fleqn,usenatbib]{mnras}

\usepackage{newtxtext,newtxmath}
\usepackage{xcolor}
\usepackage{ulem}

\usepackage[T1]{fontenc}

\DeclareRobustCommand{\VAN}[3]{#2}
\let\VANthebibliography\thebibliography
\def\thebibliography{\DeclareRobustCommand{\VAN}[3]{##3}\VANthebibliography}

\usepackage{graphicx}	
\usepackage{amsmath}	

\title[The dissociative cluster Abell 56]{Unraveling the collision scenario of the dissociative galaxy cluster Abell 56 through hydrodynamic simulations}

\author[Albuquerque et al.]{
Richards P. Albuquerque,$^{1}$\thanks{E-mail: richards\_pereira12@hotmail.com} Rubens E. G. Machado, $^{1}$ 
Rogério Monteiro-Oliveira,$^{2}$
\\
$^{1}$Departamento Acad\^emico de F\'isica, Universidade Tecnol\'ogica Federal do Paran\'a, Av. Sete de Setembro 3165, Curitiba, PR, Brazil \\
$^{2}$Institute of Astronomy and Astrophysics, Academia Sinica, Taipei 10617, Taiwan\\
\\
}

\date{Accepted 2024 April 07. Received 2024 March 11; in original form 2024 January 13}

\pubyear{2024}

\begin{document}
\label{firstpage}
\pagerange{\pageref{firstpage}--\pageref{lastpage}}
\maketitle

\begin{abstract}
In galaxy cluster collisions, the gas can be separated from dark matter halos. Abell~56 displays signatures of a dissociative bullet-like merger with a possible high inclination angle between the plane of orbit and the sky. Our objective is to provide a comprehensive description of the features observed in the collision scenario of Abell~56. Additionally, we aim to apply a potential weak lensing mass bias correction attributed to the merger to evaluate its impact on our findings. To investigate this, we perform tailored hydrodynamical $N$-body simulations, varying the impact parameter. We initially identified an early scenario at $0.12$\,Gyr after the central passage that reproduces some observational features. However, the mean temperature of $9.7$\,keV exceeded the observed value. Our best model corresponds to the late scenario at $0.52$\,Gyr after the pericenter, reproducing observed features of Abell~56, with an inclination of $58^\circ$. These features include the offset of $103$\,kpc between the main gas density peak and the south dark matter density peak, gas morphology, a line of sight relative velocity of $184$\,km\,s$^{-1}$, and a mean temperature of $6.7$\,keV. This late model provides a plausible scenario to describe the dynamics of Abell~56. The weak lensing mass bias did not significantly impact the overall dynamics of this cluster merger.
\end{abstract}




\section{Introduction}

Galaxy clusters are hierarchical structures in the universe formed by the accretion of galaxies and groups of galaxies \citep{Press74, Fakhouri10}, in a continuous process that takes place even at low redshifts \citep[e.g.,][]{Nelson23}. The collisions and subsequent minor and major mergers show signs of recent disturbances in the intracluster medium (ICM) gas, such as cold fronts and shocks \citep[e.g.,][]{Markevitch07}.

Merging systems have been detected by observing disruptions in X-ray maps, exhibiting substantial deviations from spherical symmetry in the ICM \citep{Roettiger96}. Additionally, the structural morphology of mergers in clusters can be well identified by combining observational methods of gravitational lensing and X-ray analysis \citep[e.g.,][]{Bradac08_, Monteiro-Oliveira20, Finner23, Stancioli23, Wittman23}. Another observational indication of mergers are shocks that are often detected as discontinuities in the X-ray surface brightness, but also in radio. The propagating shock waves induced by the merger may compress and amplify magnetic fields, thus accelerating relativistic electrons \citep{vanWeeren19}, usually in the periphery of the clusters. This mechanism gives rise to diffuse radio emission in the form of radio relics \citep{Feretti12} and radio haloes \citep{Cassano10}. These giant radio features are also interpreted as signatures of merging clusters.

Close pericentric collisions can lead to dissociative mergers, pulling gas out of the dark matter halos \citep{Dawson13}. Dissociative merger systems play a crucial role as they offer insights into the properties of dark matter \citep[DM; e.g.,][]{Harvey15}. The gravitational and hydrodynamics interactions in the merger process can slow down the ICM gas \citep{Molnar16}. The galaxies and the DM are practically collisionless, but the gas is not. In a near pericenter merger, most of the ICM gas may be found between the two DM components after the core passage. In the cold dark matter (CDM) scenario, DM particles will also dissociate from the gas due to their collisionless behavior. In observations of dissociative mergers, the scenario is characterized by an offset between the ICM gas and the density peaks of the matter, as is the case of the well-known `Bullet Cluster' \citep{Clowe06}. Other examples of dissociative clusters mergers include: Abell~$520$ \citep{Mahdavi07}, CI~0024+17 \citep{Jee07}, MACS~J0025.4-1222 \citep{Bradac08}, Abell~$2163$ \citep{Okabe11}, Abell~$2744$ \citep{Merten11}, DLSCL~J0916.2+2951 \citep{Dawson12}, ACT-CL~J0102-4915 \citep{Jee14}, Abell~$1758$ \citep{MonteiroOliveira17}, Abell~$3376$ \citep{Monteiro-Oliveira17b}, MACS~J$1149.5+2223$ \citep{Golovich16}, Abell~$2034$ \citep{Monteiro18}, MACS~J0417.5-1154 \citep{Pandge19}, Abell~$2256$ \citep{Breuer20}, SPT-CL~J0307-6225 \citep{Hernandez-Lang22}, and eFEDS4746/4910 \citep{Monteiro-Oliveira22}.

Dissociative mergers have been modeled through hydrodynamic $N$-body simulations to describe the collision scenario \citep[e.g.,][]{Springel07, Mastropietro08, Donnert14, Machado15, Molnar17, Lourenco20, Moura21}. Such simulations enable the exploration of the evolution of the collision even before the core passage. The plane of the sky orbits are generally simpler, in cases where they are sufficient to explain the observed features. The line of sight relative velocity provides an additional constraint in estimating the angle between the plane of the sky and the plane of the orbit. However, a more precise estimation of this angle requires the combination of observations and simulations. Numerical models, by accounting hydrodynamics process, contribute to a better understanding of this complexity.

Abell~56 (A56) is a binary dissociative cluster first reported by \cite{Abell89}. In a previous sample from the MAssive Cluster Survey (MACS) catalog, A56 was classified as a disturbed galaxy cluster \citep{Repp18}. The A56 cluster is located at $z = 0.30256 \pm 0.00058$ and the projected separation between the mass peaks was $438 \pm 206$\,kpc. In the X-ray morphology, the main peak is offset by $111 \pm 38$\,kpc from the brightest cluster galaxy (BCG) in the south. In the north cluster, it is difficult to find an associated gas peak, making the separation between the gas peak and north BCG unclear \citep{Wittman23}.

The mass of a galaxy cluster plays a crucial role in governing the kinematics of large-scale buildup. However, determining this mass becomes particularly challenging during extreme events like a merger, when the clusters are significantly perturbed from their equilibrium state. In such scenarios, weak gravitational lensing emerges as a valuable tool for estimating cluster masses, offering an advantage by not relying on assumptions of dynamical equilibrium. Nonetheless, there is a caveat. Typically, determining the cluster mass assumes that the halo follows an analytic profile, such as the Navarro–Frenk–White \citep[NFW;][]{Navarro96} model, incorporating a mass concentration relation \citep[e.g.,][]{Duffy08, Dutton14, Diemer19}. These models are often applied to describe the properties of an average halo \citep{Jing00}. Consequently, applying these approaches to individual clusters might introduce certain systematic errors, particularly in the context of merging clusters.

During the merger process, in a first-order approximation, the so-called weak lensing mass bias (hereafter WL mass bias) is modulated by variations in the cluster concentration parameter. In the extreme case of a major merger, i.e., when the mass ratio between the two most massive clusters is less than two \citep{Martel14}, the WL mass bias has the potential to overestimate real cluster masses by up to 60 percent \citep{Lee23}, suggesting that the measured masses presented in the existing literature may significantly deviate from the actual values. Beyond concerns about the accuracy of mass determination, this discrepancy can exert a substantial influence on the description of the cluster formation history.

Although multi-wavelength observations can trace various components of a cluster, including DM, gas, and stellar content, they can only capture a single moment in the cluster formation process, which spans several gigayears \citep[e.g.,][]{Okabe19,Tam20b,Monteiro-Oliveira21,Monteiro-Oliveira22a,Alden22,HyeongHan23}. While a detailed depiction of the current merger phase offers valuable insights, it alone cannot reveal the long- and short-term effects of the merger process. For instance, the anticipated subtle spatial offset between luminous and dark components, often considered an observational manifestation of self-interacting DM \citep[SIDM;][]{Adhikari22}, depends on the time elapsed after the pericentric passage, even in the late merger phases \citep{Fischer21}. Consequently, this temporal dependence must be considered to translate observed spatial offsets into a more physically meaningful quantity, such as the SIDM cross-section \citep[$\sigma/m$;][]{Wittman17}. Another time-dependent phenomenon is the impact of the cluster merger on modifying the star formation activity in member galaxies \citep[e.g.,][]{Kelkar20,Kelkar23,Wittman24}. These examples underscore the importance of numerical simulations, whether DM-only or hydrodynamical, to provide a comprehensive description of the process—from pre-collision to the late stages of cluster formation. A crucial aspect of these follow-up simulations is that they are informed by observational constraints, primarily cluster masses \citep[][]{Molnar20,Doubrawa20,Machado22,Valdarnini23}. Therefore, understanding the potential impacts of using biased masses in the merger description provided by these simulations is of utmost importance.

The main goal of this paper was to reproduce the observational features of A56 from hydrodynamic $N$-body simulations capable of reproducing its dynamic history. As a secondary goal, we also explored the application of WL mass bias correction to A56 dissociative merger scenarios based on the evolution of the concentration parameter and previous findings by \cite{Lee23}.

The paper is divided as follows. In Section~\ref{sec:simulation_setup}, we present details about the simulation setup. In Section~\ref{sec:results}, we present the results of the collision scenario, perform analyses about the evolution of the concentration parameter, and apply the WL mass bias correction to explore the consequences in the gas morphology and mean temperature. Summary and conclusions are given in Section~\ref{sec:summary_and_conclusions}. We assume a standard $\Lambda$CDM cosmology model with $\Omega_\Lambda = 0.70$, $\Omega_M = 0.30$ and $H_0 = 70$\,km\,s$^{-1}$\,Mpc$^{-1}$.

\section{Simulation setup}
\label{sec:simulation_setup}

We aim to model the A56 galaxy cluster using $N$-body hydrodynamical simulations. This cluster is composed of two substructures identified through weak lensing analysis, with the main one located to the south and the other to the north. Additionally, there is a gas peak closer to the southern part of the cluster. To replicate the dissociative collision, we constructed two structures, each consisting of a spherically symmetric halo composed of DM and gas particles. We utilized the GADGET-$4$ code \citep{Springel21}, which employs smoothed particle hydrodynamics (SPH), with a gravitational softening length of $5$\,kpc, to run the simulations. We do not consider star formation and galaxies in our analysis. The methods employed to create the initial conditions are described in detail in \cite{Rubens13} or \cite{Ruggiero17}. For the DM halo, we assumed a \cite{Hernquist90} profile:
\begin{equation}
    \rho_\mathrm{h}(r) = \frac{M_\mathrm{h}}{2 \pi} \frac{a_\mathrm{h}}{r (r + a_\mathrm{h})^3} ~,
    \label{ah}
\end{equation}
where $M_\mathrm{h}$ is the total DM mass, and $a_\mathrm{h}$ is a scale length. The
gas follows a \cite{Dehnen93} density profile:
\begin{equation}
    \rho_\mathrm{g}(r) = \frac{(3 - \gamma) M_\mathrm{g}}{4 \pi} \frac{a_\mathrm{g}}{r^\gamma (r + a_\mathrm{g})^{4 - \gamma}} ~,
    \label{ag}
\end{equation}
where $M_\mathrm{g}$ is the total gas mass and $a_\mathrm{g}$ is the gas scale length. In our simulations, we chose $\gamma = 0$, which represents a non-cool core. Once the density profile has been set, the distribution of gas temperature is determined by assuming hydrostatic equilibrium. We established a baryon fraction of $f_{\mathrm{gas}} = 0.1$, which is consistent with clusters in this mass range \citep{Lagana13}. To generate initial conditions for this work, we employed the CLUSTEP code \citep{Ruggiero17}.

To define the initial conditions, it is essential to carefully select the structural parameters of the clusters in a way that satisfies particular observational constraints. In this scenario, the A56 cluster experienced a dissociative collision, leading to potential variations in its $M_{200}$ mass over time. We chose the mass of the initial conditions in a way that ensures that the simulations match the observed data. The masses of each cluster were determined through weak-lensing analyses, resulting from the literature in $M_{200} = (4.5 \pm 0.8) \times 10^{14}$\,M$_\odot$ and $M_{200} = (2.8 \pm 0.7) \times 10^{14}$\,M$_\odot$ associated with the southern and northern galaxy subclusters respectively \citep{Wittman23}. In the simulations, each galaxy cluster contains $10^5$ particles of both DM and gas.

From the mass and redshift, the \cite{Duffy08} relation allows the estimation of the concentration parameter $c$ following the relationship:
\begin{equation}
    c = \frac{6.71}{\left ( 1 + z \right )^{0.44}} \left (\frac{M_{200}}{2 \times 10^{12} h^{-1} M_\odot} \right )^{-0.091} ~.
\end{equation}
We obtain a concentration of $c = 3.8$ and $c = 3.9$ for the south and north galaxy subclusters, respectively. The concentration parameter defined by:
\begin{equation}
    c \equiv r_{200}/r_\mathrm{s} ~,
    \label{eq:c}
\end{equation}
can be used to estimate $a_\mathrm{h}$ using the following equation \citep{Springel05}:
\begin{equation}
    a_\mathrm{h} = r_\mathrm{s} \sqrt{2[ \ln(1 + c) - c(1+c) ]} ~,
\end{equation}
where $r_\mathrm{s}$ is the scale length of the NFW profile and $r_{200}$ is the virial radius. Using the respective concentrations, we obtained $a_\mathrm{h}^\mathrm{S} = 521$\,kpc and $a_\mathrm{h}^\mathrm{N} = 434$\,kpc, 
where the index N and S indicate the north and south clusters, respectively. The parameter $a_\mathrm{g}$ can be estimated from the equation:
\begin{equation}
    a_\mathrm{g} = \frac{3 M_\mathrm{g}}{4 \pi \rho_0} ~,
    \label{eq:ag}
\end{equation}
where $M_\mathrm{g} = \frac{f_{\mathrm{gas}} M_{200}}{1 - f_{\mathrm{gas}}}$,  $\rho_0 = \mu_{\mathrm{e}} n_{\mathrm{e}} m_\mathrm{H}$ is the central gas density, the $\mu_{\mathrm{e}}$ is the mean molecular weight per electron, $n_{\mathrm{e}}$ is the electron number density, and $m_\mathrm{H}$ is the proton mass. Assuming $n_{\mathrm{e}} ^N = 3 \times 10^{-3}$\,cm$^{-3}$ and $n_{\mathrm{e}} ^S = 6 \times 10^{-3}$\,cm$^{-3}$, we obtained $a_\mathrm{g}^\mathrm{N} = 445$\,kpc and $a_\mathrm{g}^\mathrm{S} = 414$\,kpc.

After creating the initial conditions of the two structures with the parameters described above, we set up the collision scenario. The initial separation distance of each cluster was $3$\,Mpc, with an initial approach velocity of $v = 600$\,km\,s$^{-1}$. This study explored four distinct scenarios with impact parameters of $b = 0$\,kpc, $b = 200$\,kpc, $b = 400$\,kpc, and $b = 800$\,kpc.

\begin{figure*}
\begin{center}
\includegraphics[width=\textwidth]{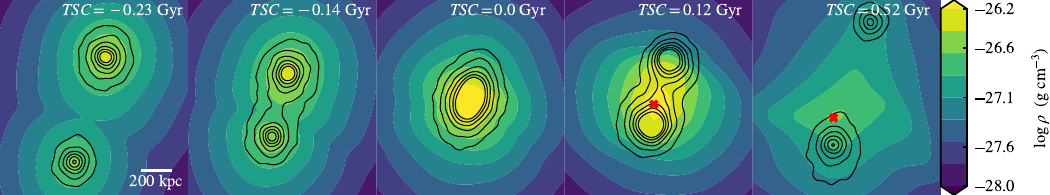}
\end{center}
\caption{Time evolution of the gas density in the southern and northern galaxy subclusters. $TSC$ (time since collision) is the time elapsed since the central passage. The optimal moments in this simulation occur at $TSC = 0.12$\,Gyr and $TSC = 0.52$\,Gyr, for the early and late scenarios, respectively, where the red cross represents the peak of the gas density after DM-gas dissociation.}
\label{fig:density_time_evolution}
\end{figure*}

\section{Results}
\label{sec:results}

After numerous simulations, we obtained the so-called `best model' that reproduces the morphology of A56. The best model represents the collision between the south and north substructures of A56,  with an impact parameter of $b = 200$\,kpc, and $v = 600$\,km\,s$^{-1}$ in the initial conditions. In this section, we will discuss the collision scenarios, the influence of the impact parameter, the evolution of the concentration, and the WL mass bias.

The main observational constraints to be satisfied by the simulations were the relative separations between the substructures. Namely, the separation between the two DM peaks must be approximately 438\,kpc; at the same time, the separation between the gas peak and the southern DM peak must be approximately 118\,kpc. These were the main quantitative criteria used to select the best snapshot in the simulation. Additionally, at least two other morphological features must be satisfied as well. First, the gas peak must lie close to the line connecting the two DM peaks. This alignment is not exactly perfect in the observations, but it is a sufficient approximation within the uncertanties in determining the peak positions. Secondly, the gas peak must be reasonably well defined, i.e. it must be sufficiently denser than its surroundings, such that it corresponds to a visually distinct concentration of gas. This criterion is necessary, because there are instances of simulations in which the dissociation has rendered the gas so diffuse, that no distinct peak can be meaningfully identified, even if the peak-finding algorithm points to one. These two qualitative morphological criteria rely on visual inspection of the snapshots, and they can be used to rule out inadequate models.

\begin{figure}
\begin{center}
\includegraphics[width=4cm]{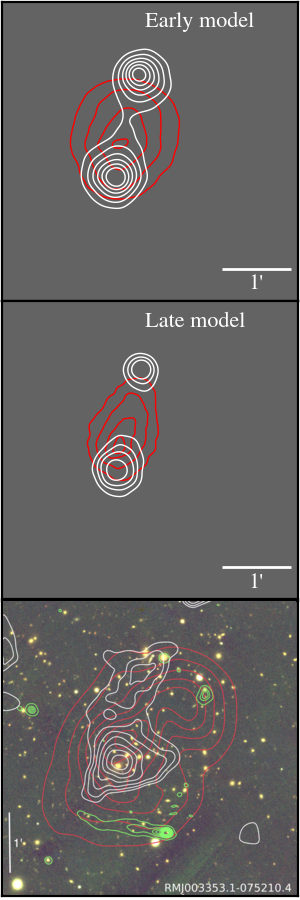}
\end{center}
\caption{The white contour lines depict the distribution of DM, while the red ones represent gas density, as simulated in the best models for the early scenario (top), and late scenario with an inclination of $i = 58 ^{\circ}$ (middle). The bottom panel displays surface mass density contours from weak lensing in white and contours from $0.4-1.25$\,keV XMM-\textit{Newton} data overlaying SDSS multiband images in red \citep[figure from][]{Wittman23}.}
\label{fig:observation_simulation}
\end{figure}

\begin{figure}
\begin{center}
\includegraphics[width=\columnwidth]{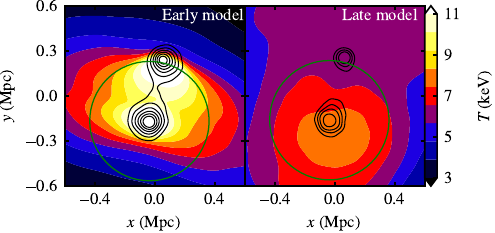}
\end{center}
\caption{Temperature map weighted by density. On the left, the early scenario with $TSC = 0.12$\,Gyr; on the right, the late scenario with an inclination of $i = 58 ^{\circ}$ and $TSC = 0.52$\,Gyr. Contour lines on the plot represent the distribution of DM, and the green circle with $r = 398$\,kpc denotes the region with mean temperatures of $9.7$\,keV (left) and $6.7$\,keV (right).}
\label{fig:temp_2x}
\end{figure}

\subsection{Collision scenarios}
\label{sec:Collision_scenarios}

Based on the observational data from the A56 cluster, we need to reproduce quantitatively a series of features. For example the offset of DM and gas peak at the point when the separation between the peaks of DM is approximately $438 \pm 206$\,kpc, and the X-ray emission morphology. Additionally, the separation between the main gas peak and the southern (northern) DM peak will be taken to be $118 \pm 41$\,kpc ($337 \pm 44$\,kpc). This is inferred from the DM-BCG and gas-BCG separations. The observational separations were obtained from the DM-gas southern offset because it is difficult to identify a gas density peak associated with the northern subcluster from X-ray emission, due to the diffuse nature of the gas in this region \citep{Wittman23}. In other words, we consider that A56 has only one relevant gas density peak -- the southern one. The secondary norteastern gas feature seen in the X-ray data will be interpreted merely as a local asymmetry, and thus disregarded from the goals of the simulations. That feature might be the result of a previous interaction, or due to the presence of additional substrucure. These are aspects that cannot be reproduced with an idealised binary collision. Fig.~\ref{fig:density_time_evolution} represents the time evolution of the gas density of A56. This figure shows the time before the pericentric passage, the pericentric passage, and the distancing of the clusters. In the two final frames, which depict the possible best instants of the simulation, the first represents the `early scenario' ($TSC = 0.12$\,Gyr), and the second represents the `late scenario' ($TSC = 0.52$\,Gyr). The best instants of the simulation were chosen based on the distance between the DM peaks and the gas offset. To obtain the projected distance between the DM peaks in the late scenario it was necessary to apply an inclination angle of $i = 58 ^\circ$, between the plane of the orbit and the plane of the sky. 

The early scenario would appear to be promising as it can reproduce a defined peak of gas in the expected position and the exact distance between the DM peaks. However, as we will see in what follows, its temperature is too high. Therefore, we propose a late scenario to mitigate this elevated temperature. In this case, a large inclination angle is needed.

Having obtained satisfactory models, we need to compare them with observational data. Fig.~\ref{fig:observation_simulation} (bottom) presents the observational configuration with the contour lines representing the DM and X-ray emission. In Fig.~\ref{fig:observation_simulation}, the top (middle) panel represents the early scenario (late scenario) of the collision. The DM peak is depicted in white, and the gas peak is shown in red. The simulated maps were rotated to match the position angle of the observations, resulting in the major cluster in the south and the minor cluster in the north. The projected separation between the DM peaks is $420$\,kpc for the early scenario and $440$\,kpc for the late scenario. In the early model, the projected distance between the south DM peak and the main gas peak is $132$\,kpc, while in the late model, it is $103$\,kpc. We conducted several binary merger simulations but were unable to reproduce the diffuse peak observed in the northwest. As far as peak separations are concerned, the models would be similarly acceptable.

\begin{figure*}
\begin{center}
\includegraphics[width=\textwidth]{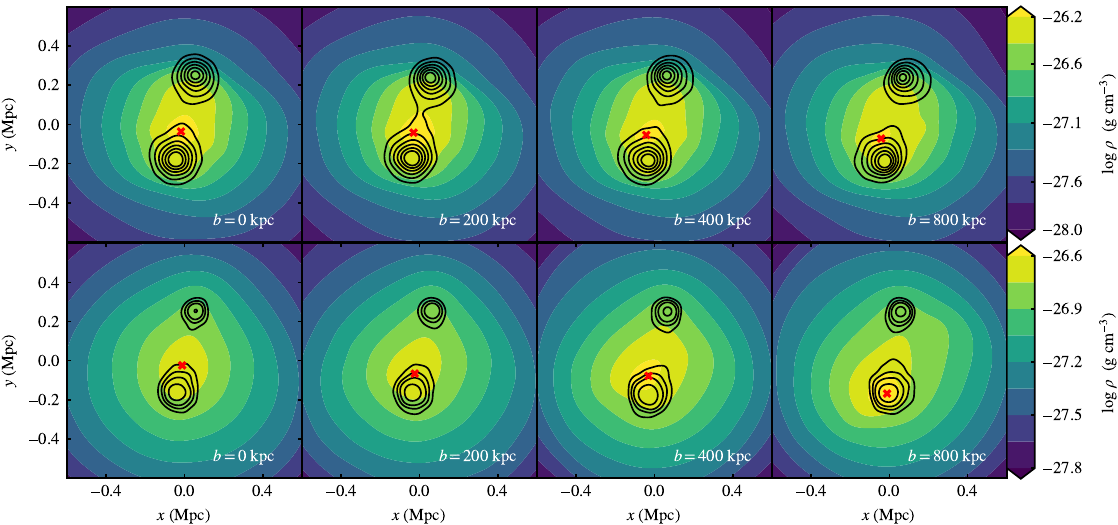}
\end{center}
\caption{Gas density map. The columns show four models with impact parameter of $0$\,kpc, $200$\,kpc, $400$\,kpc and $800$\,kpc. On the first row, we have the early scenario without inclination, while the second row represents the late scenario with inclination of $57.5 ^{\circ}$, $58 ^{\circ}$, $60 ^{\circ}$ and $58 ^{\circ}$ on the line of sight from left to right. Black contour lines on the plot represent the distribution of DM and the red cross represents the peak of the gas density.}
\label{fig:A56_density}
\end{figure*}

Regarding the temperature, there are observational constraints that need to be satisfied. This will allow us to distinguish between the early and late models.  According to \cite{Wittman23}, the temperature within a circular region with a radius of $398$\,kpc centered on the cluster is approximately $T_X = 5.9_{-0.8}^{+1.1}$\,keV. However, as we can see in Fig.~\ref{fig:temp_2x}, in the early scenario with $TSC = 0.12$\,Gyr, the estimated temperature centered in the south cluster is $9.7$\,keV. This value exceeds the observational estimate well beyond the error bar, prompting us to explore ways to mitigate this higher temperature. The late scenario is the best option to satisfy all constraints. In the late scenario, the temperature has already had time to decrease substantially. However, this results in a separation of $790$\,kpc, much larger than the observed separation. The solution involves applying a rotation of $58^\circ$, between the plane of the orbit and the plane of the sky. Therefore, the projected separation reaches the desired value. Simultaneously, the density-weighted projected temperature within a radius of $398$\,kpc corresponds to approximately $6.7$\,keV, consistent with the observational constraint.

The final temperature could, in principle, depend on the choice of certain simulation parameters. In order to attempt to decrease the temperature of the early model, we explored variations with alternative parameters. First, we considered different initial velocities, from 0 to 1200\,km\,s$^{-1}$. The result is that the final temperature maps (not shown) were not very sensitive to the choice of initial velocity. For the case with initial zero velocity, the temperatures are slightly smaller, but the contribution of the shock fronts remains dominant, such that the mean temperature in the relevant region is 9.4 keV. Thus, models with smaller velocities were not able to solve the issue of excessive temperature, and the model with 600\,km\,s$^{-1}$ was adopted as the default. Secondly, we attempted to use clusters with cool cores in the initial conditions, rather than the non-cool core clusters that were assumed throughout the paper. This was achieved by setting $\gamma=1$ for the gas density profile (equation~\ref{ag}). The steeper slope of the inner gas density profile, with the requirement of hydrostatic equilibrium, produces cool cores in the initial conditions. In the resulting cool core simulation (not shown), the central temperature was in fact somewhat decreased, but only in the region between the two DM peaks. Here again, the very hot shock fronts could not be avoided, resulting in a mean temperature of 8.2 keV within the relevant region. This result indicated that the cool core simulation could alleviate the problem, but not solve it. Thus we kept the default non-cool core models throughout the paper, and resorted to exploring the late model as the preferred candidate.

Finally, the other constraint that we aimed to satisfy in the simulation was the line of sight velocity. The line of sight velocity difference estimated by \cite{Wittman23} between the two subclusters is $\Delta v = 153 \pm 281$\,km\,s$^{-1}$. The early scenario only satisfies the velocity on the line of sight for near-zero angles, even when starting the simulation with a low approach velocity. This occurs due to the high relative velocity after the pericentric passage, with $v = 2216$\,km\,s$^{-1}$ at $TSC = 0.12$\,Gyr. With such a high velocity, even a small inclination would cause the component in the line of sight to exceed the observational upper limit. In the late model, there is sufficient time after the pericentric passage to cool the gas. Additionally, with the northern cluster located only 30 kpc short of the apocenter, the relative velocity is $v = 230$\,km\,s$^{-1}$ at $TSC = 0.52$\,Gyr. With this much lower velocity, even the application of a large inclination angle still meets the observational constraints of the line of sight velocity. Applying an inclination angle of $i = 58^\circ$ in the late scenario results in a line of sight velocity of $\Delta v = 184$\,km\,s$^{-1}$, satisfying all the main constraints and obtaining an acceptable value of the mean temperature.

\subsection{The influence of the impact parameter}
\label{sec:impact_parameter}

Here, we will investigate the effects of different impact parameters. The impact parameter is one of the least constrained properties of the collision a priori, and it could, in principle, affect the level of dissociation. All of the collision simulations presented were conducted with an initial distance of $3$\,Mpc, and the impact parameter was introduced in a perpendicular axis. For a zero impact parameter, we would have a frontal collision with a central passage. If we change the impact parameter to a nonzero value, it would induce an asymmetry in the final gas morphology. For this investigation, we conducted four different simulations with impact parameters of $0$\,kpc, $200$\,kpc, $400$\,kpc, and $800$\,kpc, as shown in Fig.~\ref{fig:A56_density}. We initiated all simulations with a low approach velocity of $600$\,km\,s$^{-1}$. Consequently, the pericentric distances were $0$\,kpc, $13$\,kpc, $27$\,kpc, and $54$\,kpc, respectively. With these low pericentric distances, the variation in impact parameters did not significantly disturb the gas morphology. This occurred because the time until the pericentric passage was long enough to align the clusters with the axis of collision, which connects the centers of the clusters. Since we added the impact parameter, the time between the beginning and the first pericentric passage varies for each simulation, and it would also interfere with the impact velocity. Therefore, we needed to incline the clusters at different angles to achieve the desired distance between the DM peaks. Hence, in the second row of Fig.~\ref{fig:A56_density} we applied inclination angles of $57.5^\circ$, $58^\circ$, $60^\circ$, and $58^\circ$ between the plane of the orbit and the plane of the sky, from the left to the right panel.

Regarding the variation in the impact parameter, there are distance constraints that need to be achieved. All the models in Fig.~\ref{fig:A56_density} are plausible, except for the late scenario with $b = 800$\,kpc, as they are consistent with the peak separation constraints, taking into account the distance error bar. We conduct the subsequent analyses of the paper with the adopted default scenario of $b = 200$\,kpc, because of a more well-defined peak of gas near the south cluster. Additionally, the distance between the peak of gas and the south cluster is $132$\,kpc for the early scenario and $103$\,kpc for the late scenario, which is close to the mean value of $118 \pm 41$\,kpc obtained from observational data in \cite{Wittman23}.  All scenarios explored in this section have in common the small pericentric distance (less than $54$\,kpc). The analysis in this subsection indicates that the model with $b = 200$\,kpc should not necessarily be regarded as the one that provides a quantitatively superior fit to the observations. Rather, it should be understood as a representative case within a family of models, all of which would be similarly acceptable within a generous margin. Nevertheless, for definiteness, the $b = 200$\,kpc model is adopted as the fiducial case.

\begin{figure}
\begin{center}
\includegraphics[width=\columnwidth]{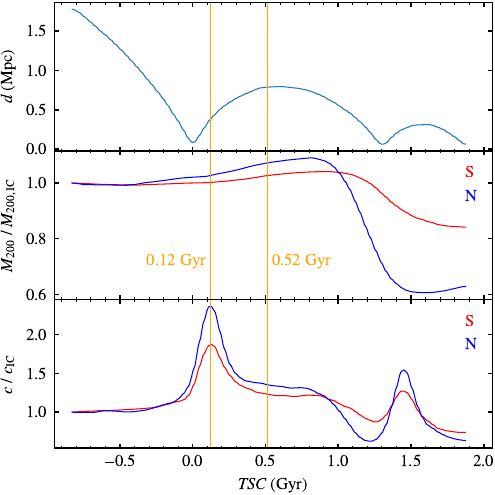}
\end{center}
\caption{At the top, we display the three-dimensional separation between the south and north clusters. In the middle, we present the time evolution of $M_{\mathrm{200}}$ normalized to its initial value. At the bottom, we depict the time evolution of the concentration parameter normalized to its initial value. The red and blue lines represent the south and north subclusters, respectively. The orange lines correspond to the early and late models.}
\label{fig:D_C_M}
\end{figure}

\subsection{Evolution of the concentration}
\label{sec:Evolution_of_the_concentration}

In this section, we will investigate the evolution of the concentration $c$ and the evolution of the virial mass $M_{200}$.  In Section~\ref{sec:Mass_bias}, the concentration parameter will be discussed in the context of the WL mass bias. Before moving to that discussion, we need to establish the properties of the default model. In Section~\ref{sec:simulation_setup}, the concentration parameter $c$ was defined in equation\,\eqref{eq:c}. To measure the time evolution of the concentration parameter, we calculated for both clusters the $r_{200}$ for each snapshot of the simulation. We considered the gas and DM particles that are part of each original cluster. Subsequently, for each snapshot, we measured the $r_{200}$ for both clusters. We treated the cluster center as identical to the location of each DM halo density peak. From the definition $\rho_{200} \equiv 200 \times \rho_\mathrm{c}$, where $\rho_\mathrm{c} = \frac{3 H^2(t)}{8 \pi G}$ is the critical density of the Universe, we identified a spherical region with a radius $r_{200}$ such that the enclosed mean density equaled $\rho_{200}$. The other parameter needed to obtain $c$, as seen in equation\,\eqref{eq:c}, is $r_{\mathrm{s}}$ from the NFW density profile. To fit the NFW density profile, we divided the cluster into spherical shells centered around the DM peak and calculated the mean density of each shell. In this way, with the mean density and mean radius of the shells, we fitted an NFW density profile and obtained the $r_\mathrm{s}$ parameter. By measuring $r_\mathrm{s}$ and $r_{200}$, we could then estimate the concentration for both the north and south clusters in every snapshot.

Fig.~\ref{fig:D_C_M} depicts, in the top panel, the three-dimensional separation between the two clusters as a function of time. Notably, the early scenario is close to the pericentric passage, while the late scenario is only $0.08$\,Gyr (or $30$\,kpc) short of the apocenter. In the middle panel of Fig.~\ref{fig:D_C_M}, the time evolution of $M_{200}$ normalized to its initial value is shown. We note that $M_{200}$ increases slowly during the first passage until a point near the second pericentric passage, where the $M_{200}$ begins to decrease. The blue line, which represents the low mass cluster, begins at $TSC \sim 0.9$\,Gyr to decrease quickly until it reaches its minimum value around the second pericentric passage. The red lines, representing the larger cluster, exhibit a similar but smoother behavior. The peak in mass is proportionally smaller, and the decline is not too abrupt. This may be because the north cluster cannot disturb the main cluster as much and does not significantly affect its structure compared to the north cluster. As noted by \cite{Lee23}, its larger concentration change in the north cluster can be caused by a more significant relative increase in gravity.

\begin{figure}
\begin{center}
\includegraphics[width=\columnwidth]{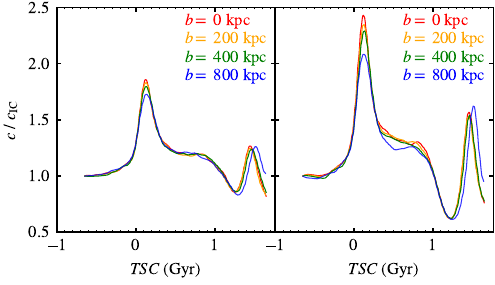}
\end{center}
\caption{The time evolution of the concentration is shown for various impact parameters. On the left, the red lines represent the south cluster, while on the right, the blue lines represent the north cluster.}
\label{fig:A56_concentration}
\end{figure}

In the initial condition, a higher concentration of DM in the cluster can result in a more compact structure, making it more resistant to disruption \citep[but see][for a more general study of how the relative DM and gas concentrations affect the dissociation]{Moura21}. In the bottom panel of Fig.~\ref{fig:D_C_M}, we compare the time evolution of the concentration, normalized to its initial value, for the south and north substructures. The best moment of the early model coincides with the concentration peaks in both clusters at $TSC = 0.12$\,Gyr. It is important to highlight that the early model was chosen based on the gas morphology and distance constraints, not the concentration. Additionally, note that the concentration peak does not occur at $TSC = 0$\,Gyr, which corresponds to the time of the pericentric passage. The concentration peak of the main (sub) cluster is $80$ per cent ($140$ per cent) higher than its initial value. For the late model, the concentration parameter of the main (sub) cluster is only $23$ per cent ($36$ per cent) higher than its initial value. Its lower value is caused by the substantial time after the pericentric passage, which is sufficient to relax the DM halo. Therefore, we noted that shortly after each pericentric passage, the halos become momentarily more concentrated. This behavior is consistent with the findings of \cite{Lee23} and \cite{Chadayammuri22}.

Despite variations in impact parameters, the gas morphology remained consistent with that of other simulated models previously presented, as we saw in Section~\ref{sec:impact_parameter}. We now analyze the evolution of the concentration across models with different impact parameters. Fig.~\ref{fig:A56_concentration} represents the time evolution of the concentration for models with $b = 0$\,kpc to $b = 800$\,kpc. Measuring both the main cluster and subcluster, we noticed that the concentration parameter peaked at its highest for $b = 0$\,kpc during the first pericentric passage, while the $b = 800$\,kpc has the lowest peak. The concentration curves for each cluster were similar even with a large impact parameter variation, and this is understandable given the similarity in pericentric distances. Regarding the south cluster, we note that the peaks and valleys were less pronounced when compared to the north cluster. Additionally, we noted that the first concentration peak decreases with the increase of the impact parameter. This behavior is also consistent with the results of \cite{Lee23} with the difference that our first concentration peak is higher, that would be because of the different mass ratios used in the simulation. 

\begin{figure}
\begin{center}
\includegraphics[width=\columnwidth]{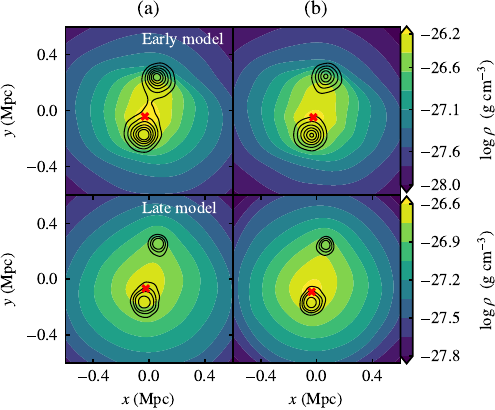}
\end{center}
\caption{Comparison between the early (top) and late (bottom) models for the gas density, showing the original mass from the weak lensing analyses (a) and the mass with bias correction (b).}
\label{fig:dens_bias}
\end{figure}

\subsection{WL mass bias}
\label{sec:Mass_bias}

The mass of the A56 cluster was estimated through weak lensing analysis by \cite{Wittman23}. Having undergone a recent collision, A56 is an unrelaxed cluster, and thus its NFW profile may not be well-fitted. The weak lensing analysis of \cite{Wittman23} was based on the mass-concentration relation from \cite{Diemer19}. This model does not consider the previous evolutionary history of the cluster. The NFW profile and the mass-concentration relation are valid only for describing the average properties of halos \citep{Jing00}. Consequently, the application of these methods to individual clusters may introduce some systematic errors, even more so in the case of merging clusters. The results of \cite{Lee23} indicate that the difference between the weak lensing and current mass can reach as much as $\sim 60$ per cent in such collisions between $10^{15}$\,M$_\odot$ halos. This mass overestimation from the weak lensing analysis is attributed to the contraction of the halo in the merger. Consequently, there is a possible WL mass bias from the A56 merger scenario. In the following, we will discuss the effects of the WL mass bias in this hydrodynamic $N$-body simulation.

In this paper, we will not analyze the weak lensings of our simulation. Our focus is on applying the maximum WL mass bias correction results from \cite{Lee23} in the A56 north and south clusters. This allows us to analyze the implications of this WL mass bias correction in an idealized hydrodynamic $N$-body simulation. In previous work, \cite{Lee23} conducted simulations to compare the virial mass and weak lensing mass over time for different mass ratio mergers. They also provided a systematic compilation of the maximum WL mass bias for each mass ratio, allowing us to consult the maximum WL mass bias for both clusters of A56. Consulting the values of the expected biases, we found that the mass for the south cluster may be overestimated by $28 \pm 12$ per cent and the north by $26 \pm 20$ per cent. We resimulated the default scenario with the WL mass bias correction to analyze the changes in our simulations. To construct the initial condition, we applied the WL mass bias correction in the $M_{200}$ for both clusters, resulting in new masses of $3.5 \times 10^{14}$\,M$_\odot$ and $2.2 \times 10^{14}$\,M$_\odot$ to the south and north clusters respectively. The $M_{\mathrm{g}}$ was calculated according to \cite{Lagana13} that estimated a baryon fraction of approximately $0.1$ for clusters across both mass ranges and the $a_\mathrm{g}$ was obtained by the equation\,\eqref{eq:ag}. Therefore, we set to the DM $a_\mathrm{h}^\mathrm{N} = 368$\,kpc and $a_\mathrm{h}^\mathrm{S} = 441$\,kpc, and to the gas $a_\mathrm{g}^\mathrm{N} = 411$\,kpc and $a_\mathrm{g}^\mathrm{S} = 381$\,kpc. With these parameters, we initialized the simulation with the same parameters as the default model: an initial separation of $3$\,Mpc, $b=200$\,kpc, and $v_0 = 600$\,km\,s$^{-1}$.

\begin{figure}
\begin{center}
\includegraphics[width=\columnwidth]{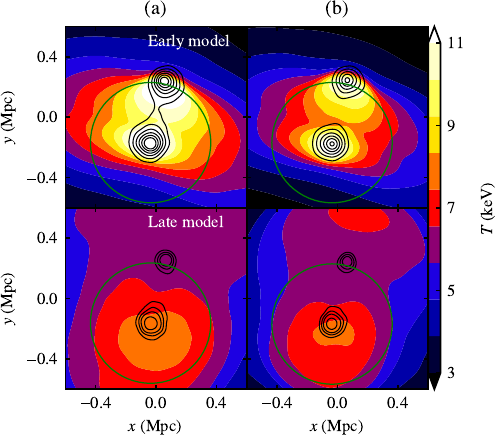}
\end{center}
\caption{Contrast between the early (upper) and late (lower) models regarding the temperature map, including the initial mass determined from weak lensing analyses (a) and the mass adjusted for bias (b). The contour lines represent the DM density and the green circle with $r = 398$\,kpc denotes the region with mean temperatures of $9.7$\,keV (top) and $6.7$\,keV (bottom) in column (a), and $8.4$\,keV (top) and $6.4$\,keV (bottom) in column (b).}
\label{fig:temp_bias}
\end{figure}

Even applying the WL mass bias correction, the simulation results satisfied the distance constraints. In Fig.\,\ref{fig:dens_bias}, the gas density morphology is presented for both cases: the original early and late scenarios in column (a), and the WL mass bias correction for each scenario in column (b). We can see that the mass peaks remain well-defined in the WL mass bias-corrected model. In the early (late) WL mass bias-corrected model, the projected distance between the DM peaks was $430$\,kpc ($440$\,kpc), and the projected distance from the south DM peak to the gas peak was $127$\,kpc ($84$\,kpc). To achieve this projected distance between the DM peaks in the late model, we applied an inclination angle of $i = 54^\circ$, between the plane of the orbit and the plane of the sky. In this scenario,  the relative velocity on the line of sight was $415$\,km\,s$^{-1}$, still within the observational error bar of $\Delta v = 153 \pm 281$\,km\,s$^{-1}$.

As we will see, the new bias-corrected model could slightly mitigate the excessive temperature of the early model. As we discussed in Section~\ref{sec:Collision_scenarios}, regarding the temperature constraint, there is a circular region with a radius of $398$\,kpc centered on the cluster is approximately $T_X = 5.9_{-0.8}^{+1.1}$\,keV. Applying the WL mass bias correction, we noted a decrease in the mean temperature, as shown in Fig.\,\ref{fig:temp_bias}. For the early (late) scenario, the temperature was $9.7$\,keV ($6.7$\,keV), and with the WL mass bias correction,  we found a mean temperature of $8.4$\,keV ($6.4$\,keV). Even with a lower mean temperature, the early scenario did not reach the expected mean temperature. For the late scenario, the mean temperature reached the observational constraint near the mean expected value of $5.9$\,keV.

\begin{figure}
\begin{center}
\includegraphics[width=\columnwidth]{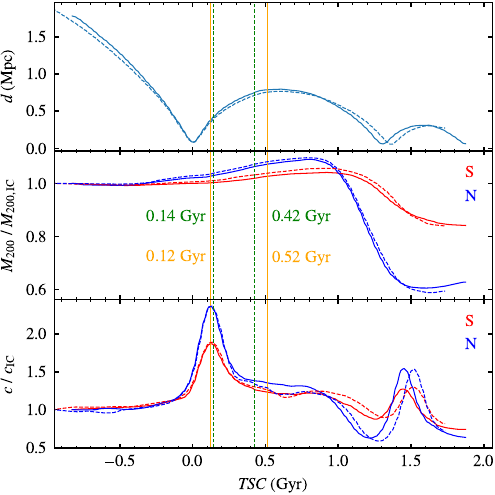}
\end{center}
\caption{The dashed lines represent the bias-corrected versions. The top panel shows the three-dimensional separation between the south and north clusters is shown. The middle panel shows, the time evolution of $M_{\mathrm{200}}$ normalized to its initial value is Shown. The bottom panel shows the evolution of the concentration parameter normalized to its initial value is shown. The red and blue lines correspond to the south and north subclusters, respectively. The orange and green lines represent the early and late models.}
\label{fig:D_C_M_bias}
\end{figure}

Since the WL mass bias correction decreases the cluster mass, the gravitational attraction is lower in this model. Consequently, dynamic differences are expected to emerge between the south and north clusters. Fig.\,\ref{fig:D_C_M_bias} in the top panel shows the three-dimensional separation between the south and north clusters. The first pericentric passage again corresponds to $TSC = 0$\,Gyr. Note that in the WL mass bias correction model, the second pericentric passage occurs approximately $0.07$\,Gyr later. The behavior of $M_{200}$ normalized to its initial value was somewhat similar in both cases, but a slight increase in $M_{200}$ normalized is noted in the WL mass bias correction model before the second pericentric passage. After the second passage, $M_{200}$ normalized reaches a lower minimum point compared to the original scenario. Even applying a correction factor of approximately $30$ per cent, it does not imply such differences in Fig.\,\ref{fig:D_C_M_bias} because the concentration and $M_{200}$ were normalized to their initial values.

Regarding the concentration parameter, up to $TSC = 0.2$\,Gyr, the normalized concentration was essentially the same in both cases, as shown in the bottom panel of Fig\,\ref{fig:D_C_M_bias}. We see that the first minimum point occurred approximately $0.06$\,Gyr later in the WL mass bias-correction model compared to the original model. Additionally, the maximum point occurred $0.08$\,Gyr later in the WL mass bias correction model. This discrepancy might be attributed to the delayed occurrence of the second pericentric passage when the WL mass bias was applied. According to \cite{Lee23}, the WL mass bias depends on the concentration. If we achieve a high concentration, the WL mass bias will also be higher. In the early scenario, at $TSC = 0.12$\,Gyr, the best moment is at the moment of maximum concentration. Consequently, the WL mass bias attains a higher value. In the late scenario, the concentration value is close to the initial one, as seen in Fig.\,\ref{fig:D_C_M_bias}. Thus, the WL mass bias is lower and falls within the error bars estimated by \cite{Wittman23}. As the early scenario did not reach the mean temperature constraints, only the late model satisfied all constraints. Consequently, in the case of A56, the application of the WL mass bias correction did not significantly alter the collision scenario, since the late model is close to the apocenter, and the concentration is nearly at its initial value.

\section{Conclusions}
\label{sec:summary_and_conclusions}

A56 is a binary galaxy cluster that underwent a dissociative merging. In our idealized simulation, we aimed to reproduce the same observational features and track the evolution of A56. Thus, we simulated distinct scenarios with different impact parameters to reach the gas morphology and the distance constraints from the observations.  The projected separation between the density peaks of DM is $438 \pm 206$\,kpc, the distance between the south cluster and the main gas density peak is $118 \pm 41$\,kpc and the line of sight difference velocity between the two subclusters is $\Delta v = 153 \pm 281$\,km\,s$^{-1}$ \citep{Wittman23}.

In a previous analysis of the MACS sample, A56 was classified as one of the most disturbed clusters among the four optical morphology classes \citep{Repp18}. Consequently, accurately reconstructing its gas morphology proved challenging in our idealized simulations. We considered two possible models in our analysis that satisfied the distance and gas morphology constraints. The early scenario depicted a plane-of-the-sky merger, reaching the expected DM peak projected separation at $0.12$\,Gyr after the pericenter. This merger resulted in a well-defined gas density peak located $132$\,kpc from the south cluster. Another crucial observational constraint is the mean temperature of $T_X = 5.9_{-0.8}^{+1.1}$\,keV, considering a circular region with a radius of $398$\,kpc centered on the cluster \citep{Wittman23}. However, the mean temperature of $9.7$\,keV in the early scenario exceeded the observational estimate well beyond the error bar. Therefore, we sought another scenario to mitigate this higher temperature. Our solution was to find a late model, close to the apocenter, at $0.52$\,Gyr after the pericenter. In this case, we obtained a well-defined projected gas density peak located at $103$\,kpc from the south cluster, and the mean temperature was approximately $6.7$\,keV, consistent with the observational constraint. Since the three-dimensional separation between the south and north clusters was $790$\,kpc, we applied a rotation of $58^\circ$ between the plane of the orbit and the plane of the sky. The early and late scenarios mentioned are part of the same simulation, representing different snapshots, with the early model shortly after the pericentric passage and the late model near the apocenter.

Notably, \cite{Wittman23} identified analogous systems in the Big Multidark Planck (BigMDPL) DM-only simulation \citep{Klypin16}. They suggested that the collision was observed relatively soon ($0.060-0.271$\,Gyr) after the pericentric passage. The observational constraints satisfied in the analogous simulated system included the projected distance between the DM density peaks, the relative velocity in the line of sight, and the masses of the north and south clusters. Since the BigMDPL simulation only includes DM, estimating the gas-DM dissociative level and the mean temperature of the cluster is not possible. Therefore, we conducted a dedicated hydrodynamic $N$-body simulation to study the gas effects and compare them with the analog from the BigMDPL simulation. Our early scenario is similar to this analog simulated one, featuring a practically plane-of-the-sky collision, a small line-of-sight relative velocity, and distance constraints. In this early model, we identified a $TSC$ of $0.12$\,Gyr and a maximum velocity at the pericentric passage that is consistent with the analog system from \cite{Wittman23}. However, we noted that in this model, the mean temperature is much higher than expected from the observations. As a result, we identified a late model that satisfies all observational features, including the mean temperature constraint, as mentioned earlier.

We explored different collision scenarios with impact parameters of $0$\,kpc, $200$\,kpc, $400$\,kpc, and $800$\,kpc. We initialized our simulations with an initial distance of $3$\,Mpc and an initial approach velocity of $600$\,km\,s$^{-1}$. With a low approach velocity, the pericentric distances reached were $0$\,kpc, $13$\,kpc, $27$\,kpc, and $54$\,kpc, respectively. Our default model was the one with an impact parameter of $200$\,kpc because of its slightly well-defined gas density peak near the south cluster. The other models are also acceptable. For a dissociative collision, a small pericentric distance was sufficient.

From the mass evolution analysis of the default model, it is evident that the $M_{200}$ changes over time. The clusters undergo compaction and experience an increase in their virial masses from the initial condition to shortly after the apocenter, due to tidal compression effects \citep{Roediger12}. After the apocenter, at $TSC \sim 0.9$\,Gyr, the $M_{200}$ starts decreasing rapidly until it reaches its minimum value near the second pericentric passage. Being the more massive cluster, the south cluster was less perturbed by the north, resulting in smoother variations in its maximum and minimum $M_{200}$ normalized to its initial value compared to the north.

Additionally, despite initiating the simulations with a default concentration based on the mass-concentration relation from \cite{Duffy08}, the concentration parameter varied throughout the simulation, reaching maximum and minimum values. In the pre-merger scenario, the concentration was close to its initial value; only at $TSC = 0.12$\,Gyr the concentration reached its peak. The early model coincides with the concentration maximum in both clusters. At this moment, the normalized concentration reached values $80$ and $140$ per cent higher than its initial value in the south and north clusters, respectively. The concentration in the late model was closer to the initial value, resulting in $23$ and $36$ per cent higher than its initial value in the south and north clusters, respectively. This lower concentration value was noted due to the substantial time after the pericentric passage. This large time was sufficient to relax the cluster.

The merger process can introduce profound changes to the internal dynamics of the clusters, altering their mass distribution considerably when compared with their undisturbed counterparts. Therefore, the mass estimation derived from the weak lensing analysis, assuming a universal mass profile, may be potentially biased in clusters undergoing a merger \citep{Euclid23, Lee23}. In our study, we re-simulated the A56 collision scenario incorporating a WL mass bias correction and evaluated the impact of this rectification on the description of the merger history, particularly in constraining the observed merger phase.

Due to the mass-concentration relation, the masses of the A56 clusters can be overestimated by $28 \pm 12$ percent for the south cluster and $26 \pm 20$ percent for the north cluster \citep{Lee23}. From the new simulation results, we found that, for the early model, the projected distance between the DM peaks was $430$\,kpc, the distance from the south DM peak to the main gas peak was $127$\,kpc, and the collision occurred near the plane of the sky. The observational constraints were satisfied in the same way they were in our first models. Since the bias-corrected masses of the clusters are lower, we expected a noticeable decrease in the mean temperature. However, the new mean temperature of the early scenario was $8.4$\,keV. The decrease of $1.4$\,keV was not sufficient to meet the expectations from the observations. Therefore, even with the WL-mass bias correction, the early model does not accurately represent the A56 cluster scenario.

The late scenario, with the maximum WL-mass bias correction, achieves a well-defined gas peak at the expected position based on observational data. Despite a high inclination of $54^\circ$ between the plane of the sky and the plane of the orbit, the relative velocity on the line of sight was $415$\,km\,s$^{-1}$, within the observational error bar. Additionally, the new mean temperature of $6.4$\,keV satisfies the observational constraint, close to the expected mean value of $5.9$\,keV. Since the WL-mass bias depends on the concentration, applying this correction to the late model of A56 did not significantly alter the collision scenario. In the late model, close to the apocenter, the concentration is near its initial value. Therefore, the cluster mass does not exceed the expected range beyond the error bar because the cluster is slightly relaxed.

\section{Summary}
The main results of our paper can be summarized as follows:

\begin{itemize}
  \item We found a late scenario, at $TSC = 0.52$\,Gyr, that successfully recovered the observational constraints of the DM peak distance and gas morphology. In this model, we identified a well-defined gas density peak near the south cluster, a mean temperature of approximately $6.7$\,keV, and a line of sight velocity of $\Delta v = 184$\,km\,s$^{-1}$, consistent with the observational constraints.

  \item The early scenario, at $TSC = 0.12$\,Gyr, achieves some of the main observational constraints, such as a well-defined density gas peak near the south cluster and the distance between the DM density peaks. However, the mean temperature of $9.7$\,keV significantly exceeds the observed value.

  \item Our best model was the late scenario, which is near the apocenter, with a concentration parameter close to the initial value. This indicates that the mass-concentration relation is practically the same as that expected by a relaxed cluster. Therefore, we conclude that in the particular configuration of the A56 merger, the WL mass bias correction did not significantly impact the collision dynamics.
\end{itemize}

In summary, we presented a specific model describing the collision scenario of the A56 merger, which successfully reproduced the observational constraints. Our best model, characterized by a high angle between the plane of the sky and the plane of the orbit, replicated the expected temperature and the distance between the density peaks. Notably, our simulations did not account for SIDM \citep[e.g.,][]{Randall08}. Future investigations should explore the inclusion of this phenomenon, particularly in dissociative scenarios like  A56, where collisions are nearly frontal. Incorporating SIDM may introduce additional dynamics, providing valuable insights into the complex nature of these merging systems.

While our findings indicate that the WL mass bias correction does not substantially alter the description of the cluster's merger history, it would be premature to consider this a generalizable result. Further investigations into various configurations, particularly exploring different mass ratios and merger phases, are essential to more reliably determine the true impact of the WL mass bias correction on the outcomes of simulations of merging clusters.

\section*{Acknowledgements}

The authors thank the anonymous referee for helpful suggestions. RPA acknowledges the financial support from UTFPR. REGM acknowledges support from the Brazilian agency \textit {Conselho Nacional de Desenvolvimento Cient\'ifico e Tecnol\'ogico} (CNPq) through grants 406908/2018-4 and 307205/2021-5, and from \textit{Funda\c c\~ao de Apoio \`a Ci\^encia, Tecnologia e Inova\c c\~ao do Paran\'a} through grant 18.148.096-3 -- NAPI \textit{Fen\^omenos Extremos do Universo}. RMO expresses gratitude to ASIAA for fostering a professional and healthy work environment.

\section*{Data Availability}

The data supporting this article will be shared upon reasonable request to the corresponding author.



\bibliographystyle{mnras}
\bibliography{A56}




\bsp	
\label{lastpage}
\end{document}